\documentclass[10pt]{article}
\usepackage{amsfonts}
\usepackage{epsfig}
\usepackage{cite}
\begin{document}
\begin{center}
{\Large\bf Nucleon Form Factors in a Light-Cone Quark Model\\
\vspace{0.3cm} %
with Two-Photon Exchange}\\%
\vspace*{1cm}
Dian-Yong Chen $^{1,2}
  \footnote{E-mail: chendy@impcas.ac.cn\ \ \ \
            Tel:  +86-931-4969308\ \ \ \
            Fax:  +86-931-4969308  }$,
Yu-Bing Dong$^{2,3}$\\
\vspace{0.3cm} %
{\ $^1$Institute of Modern Physics, \\
Chinese Academy of Sciences,\ Lanzhou,\ 730000,\ P. R. China}\\
{\ $^2$Institute of High Energy Physics, \\
Chinese Academy of Sciences,\ Beijing,\ 100049,\ P. R. China}\\
{\ $^3$ Theoretical Physics Center for Science Facilities, CAS,
Beijing 100049,\ P. R. China}
\vspace*{1cm}
\end{center}

\begin{abstract}
We estimate the two-photon exchange corrections to both proton and
neutron electromagnetic physical observables in a relativistic light
cone quark model. At a fixed $Q^2$ the corrections are found to be
small in magnitudes, but strongly dependent on the scattering angle.
Our results are comparable to those obtained from simple hadronic
model in a medium momentum transfer region.
\end{abstract}

\section{Introduction}\par

In the experimental point of view, electron-nucleon scattering is
one of time-honored tools to access the information of the intrinsic
structures of the nucleon. These structures are partly reflected by
the Sachs electric ($G_{E}(Q^2)$) and magnetic ($G_M(Q^2)$) form
factors. So far there are two experimental techniques to detect
these form factors or the form factors ratio $R=\mu_p G_{E}/G_{M}$.
The traditional one is the Rosenbluth separation method
\cite{PR-615}, which extracts the form factors ratio from the
angular dependence of the elastic electron-proton scattering cross
section. In the one-photon exchange approximation, the differential
cross section for the $e~N$ elastic scattering process is:
\begin{eqnarray}
\frac{d\sigma}{d\Omega} \propto G_{M}^2(Q^2) +\frac{\epsilon}{\tau}
G_{E}^2(Q^2),
\label{Eq-DCS-OPE}%
\end{eqnarray}
where $Q^2=-q^2$ is the momentum transfer squared, the dynamics
factor $\tau = Q^2/4M^2$ and the photon polarization parameter
$\epsilon$ relates to the laboratory scattering angle $\theta$ by
$\epsilon = (1+2(1+\tau)\tan^2\theta/2)^{-1}$. For a given value of
$Q^2$, Eq. (\ref{Eq-DCS-OPE}) shows that it is sufficient to
determine the form factors by measuring the differential cross
sections at two different values of $\epsilon$.
\par
Polarized lepton beams provide another way to access the form
factors \cite{SP-588}. In the one-photon exchange approximation, the
polarization of the recoiling proton along its motion($P_{z}$) is
proportional to $G_{M}^2$ while the component perpendicular to the
motion ($P_{x}$) is proportional to $G_{E}G_M$. It is much easier to
measure the ratios of polarizations. This method  has been used
mainly to determine the electromagnetic form factors ratio $R$
through a measurement of $P_{x}/P_{z}$ using \cite{SP-572}
\begin{eqnarray}
\frac{P_{x}}{P_{z}} = -\sqrt{\frac{2 \epsilon}{\tau (1+\epsilon)}}
\frac{G_E}{G_M}.
\label{Eq-Pol-OPE}%
\end{eqnarray}
\par%
In the framework of one-photon exchange approximation, one,
therefore, has two independent measurements of the form factors
ratio $R$. Recently, this ratio has been observed  at the Jefferson
Laboratory by the polarized method \cite{PRL-1398, PRL-092301,
PRC-055202}. It came as a surprise that the newly measured form
factors ratio is much different from the results of the Rosenbluth
separations \cite{PRD-5491,PRC-015206, PRL-142301}. As shown in Eq.
(\ref{Eq-DCS-OPE}), the form factors extracted from the Rosenbluth
separation method are strongly lie on $\epsilon-$dependent
corrections at large $Q^2$ region. After re-analyzing the next
leading order QED corrections, one finds that the two-photon
exchange (TPE) process should be restudied.
\par%
In previous literatures, there have been two different methods to
study the TPE contributions in the electron proton scattering
process. One is the simple hadronic model, in which the intermediate
states of the TPE process are taken as baryons. The known MT
corrections \cite{PR-1898,RMP-205}, which have been included in the
Rosenbluth separation method, are based on this model. It should be
emphasized that in the MT corrections, the loop integrals of TPE
contributions are evaluated by setting one of the photon's momenta
to be zero in both numerators and denominators of the amplitudes,
while the rest parts are ignored. In Ref.\cite{PRC-054320},
different from the MT corrections, the TPE contributions are
considered by neglecting one of the photon's momenta only in the
numerators of the amplitudes. Furthermore, in Ref.
\cite{PRL-142304}, the TPE contributions are evaluated by keeping
the full numerators with nucleon intermediate state. The newly
estimated results show that the corrections can at least partly
reconcile the apparent discrepancy between the two separation
methods. In the further study more intermediate states have been
taken into considerations \cite{PRL-172503,PRC-038201}.
\par%
Another approach to deal with the TPE process is the quark model. In
this model, the TPE contributions are firstly considered in the
quark level and then extended from the quark level to the baryon
level. In Ref. \cite{PRL-122301,PRD-013008}, the contributions of
TPE process have been studied at large momentum transfer and wide
scattering angle region in a parton model. In these cases, the quark
mass can be neglected. The parton model can work effectively with
large momentum transfer. However, the contributions of the TPE
process are still unknown in the medium $Q^2$ region.  In this work,
we try to calculate the TPE contributions in a light-cone quark
model and compare our results with the predictions of the simple
hadronic model. The contributions of the TPE to unpolarized
differential cross sections and polarized observables will be
evaluated at medium $Q^2$ and small $\epsilon$ regions.
\par%
This paper is organized as follows. In section \ref{Chap-EQ-TPE}, we
re-study the TPE process in quark level with massive quarks and show
a simple comparison of our results with those in Ref.
\cite{PRL-122301,PRD-013008}. In section \ref{Chap-LC-TPE}, a brief
introduction of the light-cone constituent quark model will be
addressed, and then we give our analytical expressions of the TPE
contributions in this model. The numerical results and discussions
about the TPE corrections to the differential cross sections and to
the polarized observables are displayed in section
\ref{Chap-Numerical}.
\section{Two-Photon Exchange Process in Quark Level}\par
\begin{figure}
\centering
\mbox{\epsfig{figure=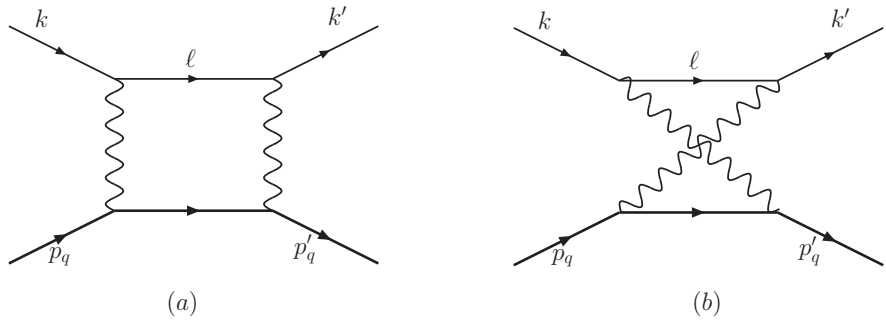,width=120mm,clip=}} %
\renewcommand{\figurename}{Fig.}
\caption{The Two-photon exchange process in quark level.}
\label{Fig-Feyn-EQ-TPE}%
\end{figure}

\label{Chap-EQ-TPE}%
As the first step, The TPE contributions in quark level (As shown in
Fig. \ref{Fig-Feyn-EQ-TPE}) will be estimated. According to parity,
time-reversal and lepton helicity conservation, the amplitude of the
TPE process in the quark level can be expanded in terms of three
independent Lorentz structures. Generally, the amplitude can be
expressed as,
\begin{eqnarray}
\mathcal{M}_{eq}^{2\gamma} = -i \frac{(e_qe)^2}{q^2}
\bar{u}(k^{\prime}) \gamma_{\mu} u(k) \bar{u}(p_q^{\prime}) \Big(
\tilde{f}_1 \gamma^{\mu} + i \tilde{f_2}
\frac{\sigma^{\mu\nu}q_{\nu}}{2m_q} +\tilde{f}_3 \frac{\gamma \cdot
K P_{q}^{\mu}}{m_{q}^2} \Big) u(p_{q}),
\label{Eq-EQ-TPE-Amp-Gen}%
\end{eqnarray}
with $P_q=(p_q+p_q^{\prime})/2$ and $K=(k+k^{\prime})/2$. Here
$\tilde{f}_i,~\{i=1,2,3\}$ are the functions of Mandelstam variables
$s^{\prime},u^{\prime},t^{\prime}$ in quark level and
\begin{eqnarray}
s^{\prime}=(k+p_{q})^2,\  u^{\prime}=(k-p_q^{\prime})^2,\
t^{\prime}=(p_{q}^{\prime}-p_{q})^2=(k-k^{\prime})^2.
\end{eqnarray}
If we ignore the electron mass, the Mandelstam variables satisfy
\begin{eqnarray}
s^{\prime} +u^{\prime} +t^{\prime} =2m_q^2.
\end{eqnarray}

\begin{figure}
\centering
\mbox{\epsfig{figure=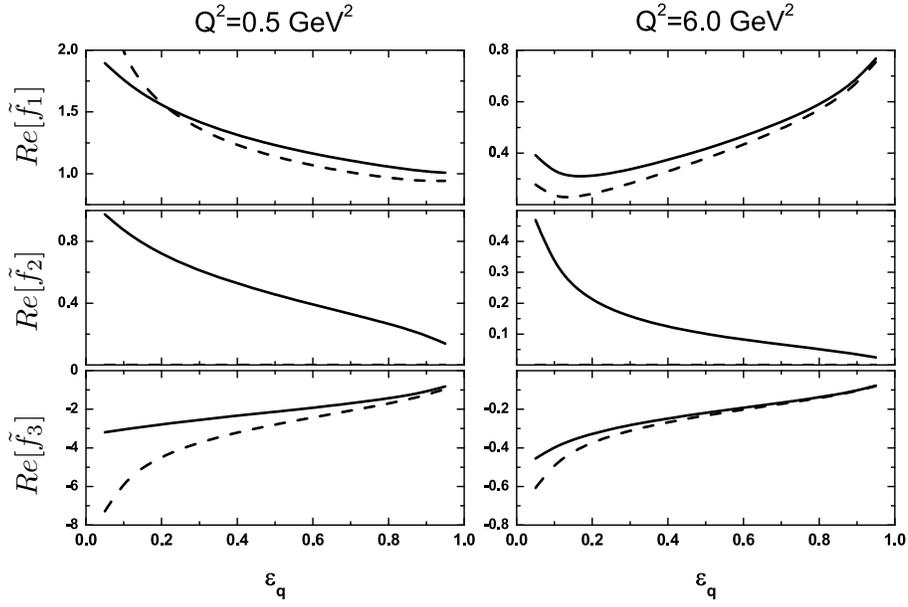,width=120mm,clip=}} %
\renewcommand{\figurename}{Fig.}
\caption{Comparisons of the real part of $\tilde{f}_i, \{i=1,2,3\}$
with massive and massless quark. Here $\epsilon_q=[(s^{\prime}-
u^{\prime})^2+ t^{\prime}(4m_q-t^{\prime})]/ [(s^{\prime}
-u^{\prime})^2 -t^{\prime}(4m_q-t^{\prime})]$. The solid curves are
the results with massive quark and the dashed curves are those with
massless quark.}
\label{Fig-fi-Comp}%
\end{figure}

In the actual calculations, the amplitudes of TPE processes
corresponding to the Feynman diagrams in Fig. \ref{Fig-Feyn-EQ-TPE}
are:
\begin{eqnarray}
\mathcal{M}_{eq (a)}^{2\gamma} &=&(ee_q)^2  \int \frac{d^4 \ell}{(2
\pi)^4} \frac{\bar{u}(k^{\prime})\gamma^{\mu} ~\hat{\ell}~
\gamma^{\nu}u(k) }{[\ell^2-m_e^2] [(k-\ell)^2-\lambda^2]
}\nonumber\\[3pt]
&&\times\frac{\bar{u}(p^{\prime}_q) \gamma_{\mu} ( \hat{p}_q
+\hat{k}-\hat{\ell}+m_q) \gamma_{\nu}
u(p_q)}{[(\ell-k^{\prime})^2-\lambda^2][(p_q+k-\ell)^2-m_q^2]}~~,
\nonumber\\[5pt]
\mathcal{M}_{eq (b)}^{2\gamma} &=&(ee_q)^2  \int \frac{d^4 \ell}{(2
\pi)^4} \frac{\bar{u}(k^{\prime})\gamma^{\mu} ~\hat{\ell}~
\gamma^{\nu}u(k) }{[\ell^2-m_e^2] [(k-\ell)^2-\lambda^2]
}\nonumber\\[3pt]
&&\times\frac{\bar{u}(p^{\prime}_q) \gamma_{\nu} (\hat{p}_q
-\hat{k}^{\prime} +\hat{\ell}+m_q) \gamma_{\mu}
u(p_q)}{[(\ell-k^{\prime})^2-\lambda^2][(p_q-k^{\prime}+\ell)^2-m_q^2]}~~,
\label{Eq-EQ-TPE-Amp}%
\end{eqnarray}
where $\hat{k} \equiv \gamma \cdot k$. The factors
$\tilde{f}_i,\{i=1,2,3\}$ in Eq. (\ref{Eq-EQ-TPE-Amp-Gen}),
therefore, can be extracted from the sum of above two amplitudes. In
Fig. \ref{Fig-fi-Comp}, we give a comparison of our results (quark
mass $m_q=0.22\ GeV$) with those in Ref. \cite{PRL-122301,
PRD-013008} in the unit of percent. In the figure the charge of the
quark is assumed as $e_q=e$. At $Q^2=6\ GeV^2$, one sees that the
real parts of our results of $\tilde{f}_1$ and $\tilde{f}_3$ are
close to the results with massless quark, especially at large
$\epsilon_q$ region. Our result about $\tilde{f}_2$ with massive
quark is comparable to $\tilde{f}_1$ and $\tilde{f}_3$ and
therefore, it should not be ignored in our calculations. Further
more, at $Q^2=0.5\ GeV^2$ the discrepancy of our results with those
in Ref. \cite{PRL-122301, PRD-013008} are even larger. Thus, one can
conclude that at large $Q^2$ region massless quark may be a good
approximation, however, at medium $Q^2$ region the quark mass is
un-neglectable. These conclusions are also suitable for the case
imaginary parts of $\tilde{f}_i, \{i=1,2,3\}$.
\par%
For further use, we separate the amplitude $\tilde{f}_1$ into soft
and hard parts, i.e. $\tilde{f}_{1}=\tilde{f}_{1}^{soft} +
\tilde{f}_{1}^{hard}$. The soft part can be obtained from Eq.
(\ref{Eq-EQ-TPE-Amp}) by neglecting one of photon momenta in the
numerators of the amplitudes. Then the soft part of $\tilde{f}_{1}$
is,
\begin{eqnarray}
\tilde{f}_1^{soft} &=&  -\frac{\alpha}{\pi} \ln \left|
\frac{s^{\prime}-m_q^2}{s^{\prime}+t^{\prime}-m_q^{2}} \right| \ln
\left|\frac{t^{\prime}}{\lambda^2}\right|.
\end{eqnarray}
One sees  $\tilde{f}_{1}^{soft}$, which is  proportional to $\ln
\lambda^2$, is IR divergent. The hard part of $\tilde{f}_{1}$ and
other structure amplitudes $\tilde{f}_2$, $\tilde{f}_3$ are IR
finite.
\par%
Here we must notice that in the parton model \cite{PRL-122301,
PRD-013008}, the soft part of $\tilde{f_1}$ is separated out by
replacing one of the photon's momenta by zero in both numerators and
denominators of the amplitudes, then one can get a three-point
Passarino-Veltman function \cite{NPB-151}, which has no analytical
representation and is much more complicate for massive quark.

\section{Two-Photon Exchange in Light-Cone Constituent Quark Model}\par

\label{Chap-LC-TPE}%
The second step of studying the TPE contributions in quark model is
to embed the amplitudes of quark level to baryon level. In parton
model \cite{PRL-122301, PRD-013008}, the general parton distribution
functions are employed. Here we perform similar calculations in a
light cone quark model.
\par%
It is well-known that the constituent quark model (CQM) developed
within a light cone framework \cite{PRL-1839,PRD-2201,AP-38,PLB-86}
appears to be an interesting tool for investigating the
electromagnetic properties of hadrons. For relativistic bound states
it provides a momentum-space Fock-state basis defined at $t^{+} =
t+z$ on the light cone, rather than the more conventional equal-time
wave functions of the instant form. On the light cone, it is
consistent to take particles on their mass shell in general. This
feature allows using light-cone spinors for quarks in multi-quark
hadron wave functions rather than propagators in the instant form.
\par%
In the light-cone quark model, for a three-quark system, the
configuration is conveniently described in terms of the
longitudinal-momentum fractions (Bjorken-Feynman variables) and
relative momentum variables:
\begin{eqnarray}
x_{j} &=& \frac{p_j^{+}}{p^{+}}, \hspace{5mm} \sum_{j=1}^{3} x_{j}
=1, \hspace{5mm} 0 \le x_j
\le 1,\nonumber\\
q_{3} &=& \frac{x_2 p_1 -x_1 p_2}{x_1 +x_2}, \nonumber\\
Q_{3} &=& (x_1 +x_2) p_3 -x_3 (p_1 +p_2) =p_3 -x_3 p_3.
\end{eqnarray}
Where $p$ and $p_{i},\ \{i=1,2,3\}$ are the momenta of nucleon and
the quarks. Here $p^{+} = \sum_{j=1}^{3} p_{j}^{+}$ reflects the
conservation of the total momentum $p$. The crucial properties of
the relative momentum variables are $Q_{3}^{+} =q_{3}^{+} =0$,
therefore they are space-like four vectors $q_{3}=-\mathbf{q}_{3
\perp}^{2}$, $Q_{3}^{2} =-\mathbf{Q}_{3 \perp}^2$. These six
relative variables $x_1$, $x_2$, $\mathbf{q}_{3 \perp}$,
$\mathbf{Q}_{3 \perp}$ are translational invariant and invariant
under the three light-cone boost \cite{PRD-768}.
\par%
In present work, the calculations are performed in a symmetric frame
\cite{PLB-204,EPJC-409}, which are the same as those in the parton
model \cite{PRL-122301, PRD-013008}. In such a frame the Mandelstam
variables in baryon level are
\begin{eqnarray}
s &=& (p+k)^2=-\frac{1+\eta}{4\eta}t +(1+\eta) M^2,\nonumber\\
u &=& (p-k^{\prime})^2=\frac{1-\eta}{4\eta}t +(1-\eta) M^2, \nonumber\\
t &=& (k-k^{\prime})^2= (p^{\prime}-p)^2,
\end{eqnarray}
with $\eta = (s-u-2\sqrt{M^4-su})/(4M^2-t)$. Moreover, in above
frame, we have a large $p^{+}$ (the $'+'$ component of the initial
proton), then the transverse momenta of the spectator quarks are
supposed to be small relative to $p^{+}$ and can be neglected. Based
on such approximation, the Mandelstam variables of the quark level
can directly connect to those of the baryon level by
\begin{eqnarray}
s' &=& -\frac{(x+ \eta)^2}{4 x \eta} t + \frac{x + \eta}{x}
m_q^2,\nonumber\\[3pt]
u' &=& \frac{(x- \eta)^2}{4 x \eta} t + \frac{x - \eta}{x} m_q^2,
\end{eqnarray}
where $x=p_q^{+}/p^{+}$ is the ratio of $'+'$ component of the
active quark (quark interacting with the external field) momentum
and nucleon momentum.
\par%
The symmetric frame can be taken as a special Drell-Yan frame with
the essential feature $q^{+}=0$ \cite{PRD-2201, PRL-181}. In such a
frame, the form factors $F_{1}$ and $F_{2}$ under one-photon
exchange approximation can be determined from the $J^{+}$ matrix
elements alone, i.e.,
\begin{eqnarray}
e F_{1}(q^2) &=& \frac{M}{P^{+}} \langle N(P^{\prime})\uparrow
|J^{+}| N(P)\uparrow \rangle\ \ ,\nonumber\\[5pt]
\frac{q_{L}}{2M} e F_{2}(q^2) &=& -\frac{M}{P^{+}} \langle
N(P^{\prime})\uparrow |J^{+}| N(P)\downarrow \rangle\ \ .
\label{Eq-FFs-1g}
\end{eqnarray}
The matrix element for the three-quark nucleon wave function
$\psi_N$ reads,
\begin{eqnarray}
\langle N \lambda^{\prime} |\frac{J^{+}}{P^{+}}| N \lambda \rangle =
\sum_{j=1}^{3} \int d\Gamma \ \psi_{N}^{\dag} (x_{i}^{\prime},
q_{3}^{\prime}, Q_{3}^{\prime}, \lambda^{\prime}) \frac{ J^{+}_{q}
}{\sqrt{p_{j}^{\prime+}p_{j}^{+}}} \psi_{N} (x_i,q_3, Q_3, \lambda),
\label{Eq-PHV}
\end{eqnarray}
with the invariant phase-space volume element
\begin{eqnarray}
d\Gamma =\frac{1}{(2\pi)^6} d^2 \mathbf{q}_{3\perp} d^2
\mathbf{Q}_{3\perp} \delta \Big(\sum_{i=1}^{3} x_{i} -1\Big)
\prod_{i=1}^{3} \frac{dx_{i}}{x_{i}},
\end{eqnarray}
and $\lambda$ denotes the spin of nucleon. $J_{q}^{\mu}$ is the
electromagnetic current of the active quark with charge $e_{j}$. In
the one-photon approximation, we have,
\begin{eqnarray}
J_{q \mu}^{1\gamma} =e_j \bar{u}(p_j^{\prime}) \gamma^{\mu} u(p_j)
\label{Eq-EQC-1g}%
\end{eqnarray}
After considering TPE contributions, the electromagnetic current of
the active quark can be derived from Eq. (\ref{Eq-EQ-TPE-Amp-Gen}),
it is,
\begin{eqnarray}
J_{q \mu}^{2 \gamma} =e(e_j)^2 \bar{u}(p_j^{\prime}) \Big(
\tilde{f}_1 \gamma^{\mu} + i \tilde{f_2}
\frac{\sigma^{\mu\nu}q_{\nu}}{2m_q} +\tilde{f}_3 \frac{\gamma \cdot
K P_{q}^{\mu}}{m_{q}^2} \Big) u(p_{j}).
\label{Eq-EQC-2g}%
\end{eqnarray}
Meanwhile, in the baryon level, after including the TPE
contributions, a new term will be introduced in the nucleon
electromagnetic vertex, and the vertex becomes,
\begin{eqnarray}
\Gamma^{\mu}=\tilde{F}_1\gamma^{\mu} +\tilde{F}_2
\frac{i\sigma^{\mu\nu}q_{\nu}}{2M} + \tilde{F}_3 \frac{\gamma \cdot
K P^{\mu}}{M^2}.
\end{eqnarray}
Then, after taking the TPE processes into considerations, the
corresponding matrix elements in Eq. (\ref{Eq-FFs-1g}) become
\begin{eqnarray}
\frac{M}{P^{+}} \langle N(P')\uparrow|J_{tot}^{+}|N(P) \uparrow
\rangle &=& e (\tilde{F}_1+\frac{1}{2} (\eta -\frac{q_{L}^2
(\eta^2 -1)}{4 m^2 \eta}) \tilde{F}_{3}) , \nonumber\\
\frac{M}{P^{+}} \langle N(P') \uparrow| J_{tot}^{+}|N(P) \downarrow
\rangle &=& -\frac{q_L}{2M}e (\tilde{F}_{2}-\eta \tilde{F}_{3} ),
 \label{Eq-FFs-2g}
\end{eqnarray}
where $J^{tot}_{\mu} =J_{q\mu}^{1\gamma} + J_{q\mu}^{2\gamma}$. With
the TPE current of the active quark and the definitions in  Eq.
(\ref{Eq-PHV}) we can get some information about the TPE corrections
to nucleon form factors.

\section{Numerical Results and Discussions}\par
\label{Chap-Numerical}%

In this work no more parameters are needed than those in the
one-photon approximation \cite{PRL-1839, PRD-2201, AP-38, PLB-86}.
The nucleon wave function $\psi_N$ is quoted from Ref.
\cite{PRD-2201,PLB-86}, in which the quark mass is set as $m_q=0.22\
GeV$ and a gaussian form wave function is employed with a parameter
$\beta=0.55\ GeV$ \cite{PRD-2201}.
\par%
From Eq. (\ref{Eq-FFs-2g}), we can easily get the TPE contributions
to the electromagnetic current matrix element. After considering TPE
process, there are three independent Lorentz structures in the
nucleon electromagnetic vertex, so we can not separate out all the
information of the TPE corrections to the nucleon form factors from
the two identities in Eq. (\ref{Eq-FFs-2g}). As we know, the nucleon
electric form factors are smaller than the magnetic form factors,
especially for the neutron, so one can suppose TPE corrections to
$G_{E}^{N}$ is zero \cite{PRC-015202}, namely $\Delta G_{E}=0$. With
this assumption, we can estimate the TPE effect on the nucleon form
factors. In our calculations, we define $Y^{D}_{2\gamma} = \nu
\tilde{F}_3/M^2G_D$ instead of $Y_{2\gamma} = \nu \tilde{F}_3/M^2
G_M$ with $G_{D}$ being the form factor in dipole form. In this way,
we can represent the TPE corrections without considering the results
under one-photon approximation. In order to make our calculations to
be comparable to the experimental data, we have to consider the IR
divergent part in $\tilde{f}_1$, i.e. $\tilde{f}_{1}^{soft}$,
separately. Similar tricks can be done as those in parton model
\cite{PRL-122301,PRD-013008}. The soft parts are evaluated in a
simple hadronic model with the nucleon as the intermediate state,
while the contributions of the hard parts, including
$\tilde{f}_1^{hard}$, $\tilde{f}_2$ and $\tilde{f}_3$, are estimate
from the matrix elements as shown in the left hand of Eq.
(\ref{Eq-FFs-2g}).
\par
\begin{figure}
\centering
\mbox{\epsfig{figure=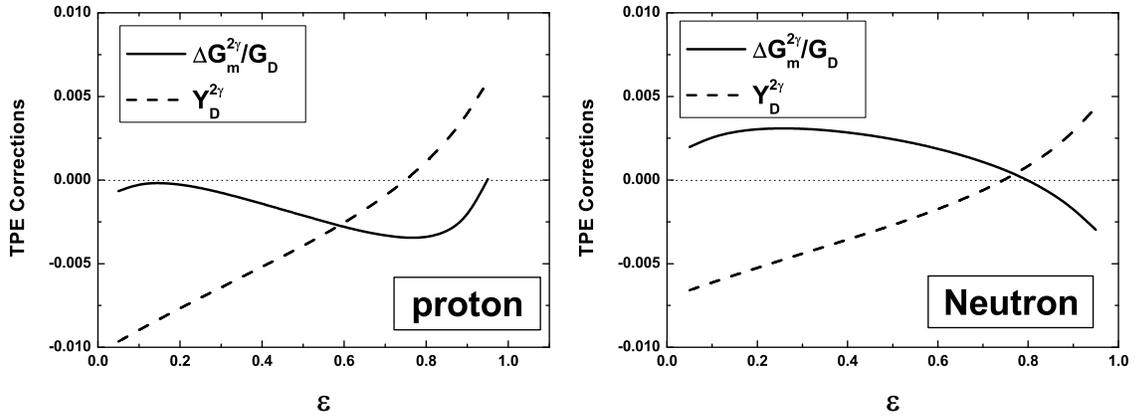,width=150mm,clip=}} %
\renewcommand{\figurename}{Fig.}
\caption{Hard part of two-photon exchange contributions to nucleon
form factors. The solid lines are the corrections to magnetic form
factors and the dashed lines are those for $Y_{D}^{2\gamma}$.}
\label{Fig-DFFs}%
\end{figure}

%
In Fig. \ref{Fig-DFFs}, we show our results for the hard part of TPE
corrections to nucleon form factors at $Q^2=1\ GeV^2$ with the
assumption $\Delta G_{E}=0$. The right (left) panel is the results
about the proton (neutron). For the TPE corrections to the nucleon
form factors, one can find, their magnitudes are very small, but
these corrections are strongly dependent on the photon polarization
parameter $\epsilon$. These features are similar to the conclusions
drawn from the calculations in parton model
\cite{PRL-122301,PRD-013008} and simple hadronic model
\cite{PRC-034612}. Since in present work we can not separate all the
TPE corrections to nucleon form factors and moreover, as we
mentioned in section \ref{Chap-EQ-TPE}, the soft part separated from
$\tilde{f}_1$ is not the same as those in parton model, then in
baryon level, our results about the hard part of TPE corrections to
form factors are not exactly comparable to those obtained in the
literatures \cite{PRL-122301,PRD-013008,PRC-034612}.
\par%
After considering TPE corrections, the total unpolarized
differential cross section is
\begin{eqnarray}
\frac{d \sigma^{t}}{d \Omega} \equiv \frac{d \sigma^{1 \gamma}}{d
\Omega} (1 + \delta^{2 \gamma}) = \frac{d \sigma^{1 \gamma}}{d
\Omega} + \Big(\frac{d \sigma^{2\gamma}_{soft}}{d \Omega}-\frac{d
\sigma^{2\gamma}_{MT}}{d \Omega} \Big) +\frac{d
\sigma^{2\gamma}_{hard}}{d \Omega}  .
\end{eqnarray}
The subscript 'soft' and 'hard' denote the soft part and the hard
part of TPE corrections respectively, while 'MT' means the MT
corrections which have been included in the experimental data. As we
have mentioned above, the soft part of the TPE corrections is
evaluated in simple hadronic model, while the hard part can be
evaluated as
\begin{eqnarray}
\frac{d \sigma^{2\gamma}_{hard}}{d \Omega} = 2 G_{M}^{1 \gamma} Re
\Big[ \Delta G_M +\epsilon \frac{\nu}{M^2} \tilde{F}_3 \Big] +2
\frac{\epsilon}{\tau} G_E^{1\gamma} Re \Big[ \Delta G_E + G_{D} Y_{2
\gamma}^{D} \Big].
\end{eqnarray}
With the assumption $\Delta G_E =0$ and the results of TPE
corrections to other form factors we can get the hard part
corrections to the cross section. Here the form factors under
one-photon approximation $G_{E,M}^{1\gamma}$ are taken from the
Rosenbluth experimental data \cite{PRD-5671, PLB-26, PRL-122002}.
\par
\begin{figure}
\centering
\mbox{\epsfig{figure=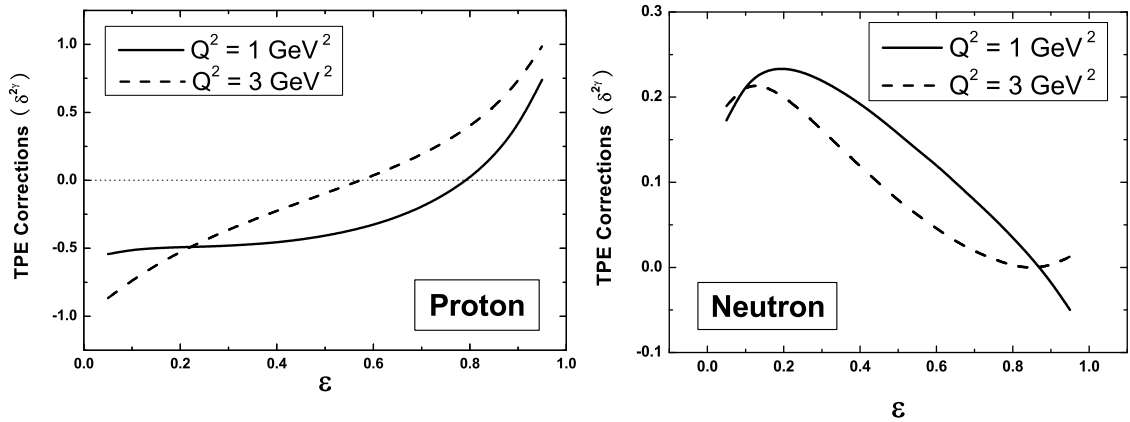,width=150mm,clip=}} %
\renewcommand{\figurename}{Fig.}
\caption{Two-photon exchange contributions to unpolarized
differential cross sections. The solid lines are the results at
$Q^2=1\ GeV^2$, while the dashed lines are those at $Q^2= 3\
GeV^2$.}
\label{Fig-DCS}%
\end{figure}

%
The TPE corrections to nucleon unpolarized differential cross
sections are presente in Fig. \ref{Fig-DCS}. The right (left) panel
shows the TPE corrections to proton (neutron) differential cross
sections. For proton case, the TPE corrections are about $1.3 \%$ at
$Q^2= 1 \ GeV^2$ and nearly $2\%$ at $Q^2=3 \ GeV^2$. That means,
with $Q^2$ increasing, the TPE correction increases too, which is
consistent with the conclusion drawn from parton model and simple
hadronic model. However, in the simple hadronic model, when only
considering nucleon as the intermediate state \cite{PRC-034612}, the
TPE corrections to differential cross sections are about $2 \%$ and
$4\%$ at $Q^2=1 \ GeV^2$ and $Q^2= 3 \ GeV^2$ separately. While
including $\Delta(1232)$ as well as the nucleon in the calculations,
the TPE corrections to proton differential cross sections are about
$1.8 \%$ and $2.8 \%$ at above two momentum transfer points. Then we
can conclude that our results about TPE contributions to proton
differential cross sections are comparable to those in simple
hadronic model.

For neutron, our results are rather small, about $0.3 \%$ and $0.2
\%$ at $Q^2=1 \ GeV^2$ and $Q^2= 3 \ GeV^2$ separately, which are
far less than $0.8 \%$ and $1.5 \%$ in simple hadronic model
\cite{PRC-034612}. In Ref. \cite{PRC-034612}, only nucleon is
considered as the intermediate state. When  more nucleon resonances,
such as $\Delta(1232)$, are included, the TPE corrections to neutron
differential cross sections are also supposed to be weaken as the
case of proton \cite{PRL-172503}. It should be reiterated that in
our present work we can not separate out all the TPE corrections to
form factors exactly, and therefore, we can just give a rough
estimate about corrections to the differential cross section.
\par
\begin{figure}[htb]
\centering
\mbox{\epsfig{figure=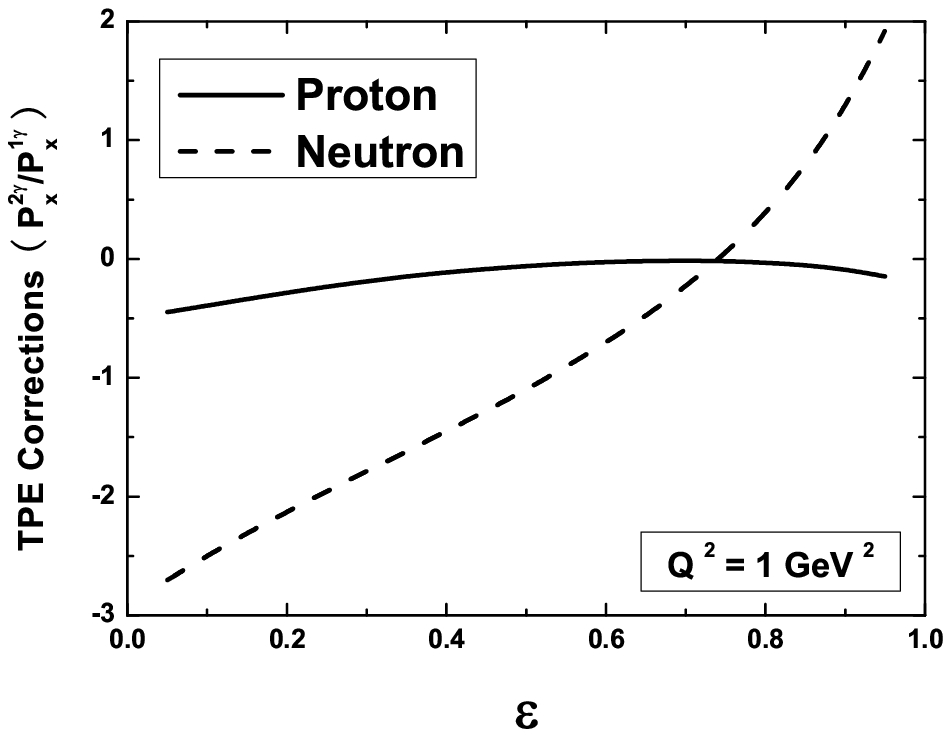,width=75mm,clip=}} %
\mbox{\epsfig{figure=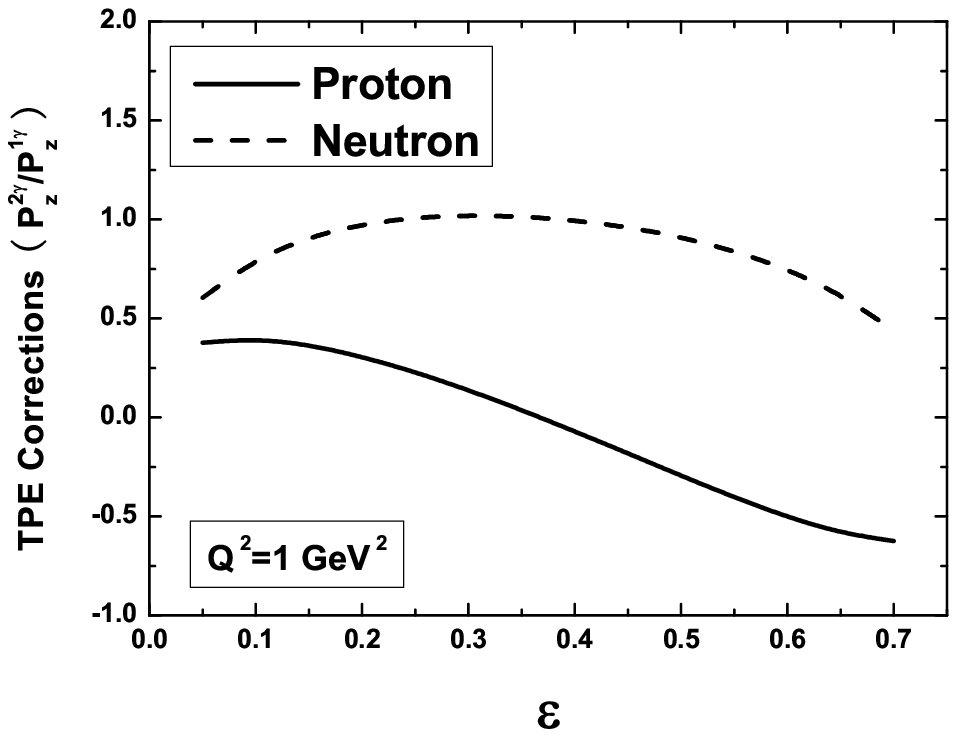,width=75mm,clip=}} %
\renewcommand{\figurename}{Fig.}
\caption{Two-photon exchange contributions to polarized observables
at $Q^2 =1\ GeV^2$. The solid lines are results for proton and the
dashed lines are those for neutron.}
\label{Fig-Dpol}%
\end{figure}

%
The polarized observables $P_x$ and $P_z$ have extra terms after
considering TPE corrections. They are expressed as:
\begin{eqnarray}
P_{x} &=& -\sqrt{\frac{2 \epsilon (1- \epsilon)}{\tau}} \Big(
\frac{d\sigma^{un}}{d\Omega} \Big)^{-1} \Bigg\{ G_{E} G_M + \Big[
G_{E} \Delta G_M + G_M \Delta G_E + G_M G_D Y_{2\gamma}^{D} \Big]
\Bigg\},\nonumber\\[3pt]
P_{z} &=& \sqrt{1- \epsilon^2} \Big( \frac{d\sigma^{un}}{d\Omega}
\Big)^{-1} \Bigg\{ G_{M}^2 + 2 \Big[ G_{M} \Delta G_M
+\frac{\epsilon}{1- \epsilon} G_M G_D Y_{2\gamma}^{D} \Big] \Bigg\}.
\end{eqnarray}
In above expression, the terms in the square brackets are the
contributions from TPE, i.e. $P_{x,z}^{2 \gamma}$. After taking
$\Delta G_{E, M} =0$ and $Y_{2 \gamma}^{D}=0$, the above expression
will be reduced to those under one-photon exchange approximation ,
i.e. $P_{x,z}^{1 \gamma}$. As shown in Fig . \ref{Fig-Dpol}, we give
the results for TPE corrections to the polarized observables at
$Q^2=1 GeV^2$. The left panel shows the ratios of $P_{x}^{2\gamma}$
and $P_{x}^{1\gamma}$ in the unit of percent. For proton, the TPE
correction is about $0.5\%$, which is close to the results in simple
hadronic model \cite{PRC-034612}. But for neutron, the correction is
much larger, is about $5 \%$. In one photon approximation,
$P_x^{1\gamma}$ is proportional $G_{E} G_{M}$ and $G_{E}^{n}$ is so
small that $P_{x}^{1\gamma}$ is rather tiny. However, the TPE
contributions contain term $G_{M} G_{D} Y_{2\gamma}^{D}$ and
$G_{M}^{n}$ is much larger than $G_{E}^{n}$, then the ratio
$P_{x}^{1\gamma}/P_{x}^{2\gamma}$, for the neutron may be relative
large. The right panel in Fig. \ref{Fig-Dpol} shows the results for
the TPE corrections to $P_z$. For proton, the correction is about $1
\%$, which is about two times of the results in simple hadronic
model \cite{PRC-034612}. For neutron, our results is about $0.5 \%$.
The TPE corrections to neutron polarized observables keep unknown in
medium $Q^2$ region in both simple hadronic model and parton model.
\par%
To summarize, we have studied the TPE corrections in a relativistic
constituent quark model for the first time. The quark mass is found
to be un-neglectable in medium momentum transfer region. In present
work, we separate out the TPE contributions to form factors with the
assumption $\Delta G_{E} =0$, and then study the TPE contributions
to the differential cross sections as well as to the polarized
observables. It is found that the TPE corrections to electron-proton
scattering differential cross sections are rather small in magnitude
but with strong $\epsilon$ dependence, and the TPE corrections
become important at high $Q^2$ region. These conclusions are
consistent with those drawn from parton model and simple hadronic
model. However, For neutron, our results are much smaller than the
results in hadronic model. This can be interpreted as not exactly
extracting the TPE contributions to the form factors in our
calculations as well as not including more nucleon resonances in the
hadronic model. For polarized observables, the results for proton in
present work are a little larger, but still comparable to those in
hadronic model. Further more, there are some uncertainties for the
results in hadronic model because it keeps unknown that how the
nucleon resonances, especially $\Delta(1232)$, effect the TPE
contributions to polarized observables. For the case of neutron, the
TPE corrections to polarized observables in such low momentum
transfer region have not been studied in previous literatures.
\par%
In our calculations, we also try to suppose the TPE contributions to
other form factors to be zero and then study TPE contributions to
electromagnetic physical observables. We can get similar results for
proton, but different results for neutron due to the electric form
factor of neutron is far less than the magnetic form factor. In
principle, the TPE corrections can be separated by Eq.
(\ref{Eq-FFs-2g}) together with the matrix element of other
component, such as the matrix element of $J^y$. Unfortunately, for
the contributions of zero mode\cite{PRD-116001-113007}, those
separation may be much complicate and we believe that it can be
evaluated in a di-quark model with light-cone formulism. This work
is under process.

\section{Acknowledgments}
\par\noindent\par\noindent\par
This work is supported  by the NSF of China under Grant No.
10775148, by CAS Grant No. KJCX3-SYW-N2 (YBD), and in part by the
National Research Council of Thailand through Suranaree University
of Technology and the Commission of High Education, Thailand.
Discussions with Yu-Chun Chen and Hai-Qing Zhou are appreciated.

\clearpage\newpage

\clearpage\newpage

Fig. 1: The Two-photon exchange process in quark level \\

Fig. 2: Comparisons of the real part of $\tilde{f}_i, \{i=1,2,3\}$
with massive and massless quark. Here $\epsilon_q=[(s^{\prime}-
u^{\prime})^2+ t^{\prime}(4m_q-t^{\prime})]/ [(s^{\prime}
-u^{\prime})^2 -t^{\prime}(4m_q-t^{\prime})]$. The solid curves are
the results with massive quark and the dashed curves are those with
massless quark. \\

Fig. 3: Hard part of two-photon exchange contributions to nucleon
form factors. The solid lines are the corrections to magnetic form
factors and the dashed lines are those for $Y_{D}^{2\gamma}$. \\

Fig. 4: Two-photon exchange contributions to unpolarized
differential cross sections. The solid lines are the results at
$Q^2=1\ GeV^2$, while the dashed lines are those at $Q^2= 3\
GeV^2$.\\

Fig. 5: Two-photon exchange contributions to polarized observables
at $Q^2 =1\ GeV^2$. The solid lines are results for proton and the
dashed lines are those for neutron.

\clearpage\newpage%
\clearpage\newpage%
\clearpage\newpage%
\clearpage\newpage%
\clearpage\newpage%

\end{document}